\documentstyle[11pt]{article}

\newcommand{\sect}[1]{\setcounter{equation}{0}\section{#1}}
\newcommand{\subsect}[1]{\subsection{#1}}

\newcommand{\vs}[1]{\rule[- #1 mm]{0mm}{#1 mm}}

\newcommand{\lbl}[1]{\label{eq:#1}}
\newcommand{\rf}[1]{(\ref{eq:#1})}
\newcommand{\nn}{\nonumber}
\newcommand{\be}{\vs{2}\begin{equation}}
\newcommand{\ee}{\vs{2}\end{equation}}

\newcommand{\bea}{\begin{eqnarray}}
\newcommand{\ena}{\end{eqnarray}}
\newcommand{\nnbea}{\begin{eqnarray*}}
\newcommand{\nnena}{\end{eqnarray*}}

\newcommand{\lra}{\ \longrightarrow\ }

\newcommand{\ovl}[1]{\overline{#1}}

\newcommand{\dps}{\displaystyle}

\newcommand{\bz}{{\ovl{z}}}
\newcommand{\bc}{{\ovl{c}}}

\newcommand{\bZ}{{\ovl{Z}}}
\newcommand{\yz}{y_{z}}
\newcommand{\ybz}{\ovl{y}_{\bz}}

\newcommand{\YZ}{Y_{Z}}

\newcommand{\YbZ}{\ovl{Y}_{\bZ}}

\newcommand{\zbz}{(z,\bz)}
\newcommand{\zy}{{(z,y)}}

\newcommand{\zY}{{(z,Y)}}
\newcommand{\ZY}{{(Z,Y)}}
\newcommand{\Zn}{Z^{(n)}}

\newcommand{\YbY}{(\YZ,\YbZ)}

\newcommand{\cW}{{\cal W }}
\newcommand{\cU}{{\cal U }}
\newcommand{\sS}{{\cal S }}
\newcommand{\cK}{{\cal K }}
\newcommand{\cF}{{\cal F }}
\newcommand{\cC}{{\cal C }}

\newcommand{\cD}{{\cal D }}

\newcommand{\cbD}{\ovl{\cal D }}
\newcommand{\cT}{{\cal T }}

\newcommand{\cS}{{\cal S }}

\newcommand{\cX}{{\cal X }}

\newcommand{\cJ}{{\cal J }}

\newcommand{\prt}{\partial}
\newcommand{\prtz}{\partial_z}
\newcommand{\prtbz}{\ovl{\partial}_{\ovl{z}}}
\newcommand{\lambdan}{\lambda_z^{\Zn}}
\newcommand{\lambdaZ}{\lambda_z^{Z}}
\newcommand{\lambdadue}{\lambda_z^{Z^{(2)}}}
\newcommand{\lambdapj}{\lambda_z^{Z^{(p_j)}}}
\newcommand{\prtZ}{{\partial}_{ Z}}

\newcommand{\prtY}{\frac{\partial}{\partial Y_Z}}
\newcommand{\prtbY}{\frac{\partial}{\partial \ovl{Y}_{\bZ}}}
\newcommand{\bprt}{\ovl{\partial}}
\newcommand{\blambda}{\ovl{\lambda}}
\newcommand{\mub}{\ovl{\mu}}

\newcommand{\ZBZ}{(Z,\ovl{Z})}
\newcommand{\ZBZn}{(Z^{(n)},\ovl{Z}^{(n)})}
\newcommand{\ZBZdue}{(Z^{(2)},\ovl{Z}^{(2)})}

\newcommand{\zbzp}{(z',\ovl{z}')}

\begin{document}
\renewcommand{\thefootnote}{\fnsymbol{footnote}}
\setcounter{page}{1}

\title{\bf $W$-algebras from symplectomorphisms}

\vskip 1.5cm

\author{G. BANDELLONI $^a$
\hskip 0.2mm and S. LAZZARINI $^b$\\[6mm]$^a$ Dipartimento di Fisica dell'
Universit\'a di Genova\\
and \\
Istituto Nazionale di Fisica Nucleare, INFN, Sezione di Genova\\
via Dodecaneso 33, I-16146 GENOVA Italy\\
e-mail : {\tt beppe@genova.infn.it}\\[4mm]
$^b$ CPT-CNRS Luminy, Case 907,F-13288 MARSEILLE Cedex, France\\
e-mail : {\tt sel@cpt.univ-mrs.fr} }

\maketitle

\vskip 1.5cm

\begin{abstract}
It is shown how $W$-algebras emerge from very peculiar canonical 
transformations with respect to the canonical symplectic  
structure on a compact Riemann surface.
The action of smooth diffeomorphisms of the cotangent bundle
on suitable generating functions  
is written in the BRS framework while a $W$-symmetry 
is exhibited. Subsequently, the complex structure of the symmetry spaces
is studied and the related BRS properties are discussed.
The specific example of the so-called $W_3$-algebra is treated  
in relation to some other different approaches.
\end{abstract}

\newpage

\sect{Introduction}

\indent

In the last decade, a large body of literature has been devoted to
the study of the so-called $W$-algebras.  The latter were first
introduced as higher spin extension of the Virasoro algebra,
\cite{Zam1,Zam2}, through the operator product expansion (OPE) of
the stress-energy tensor and primary fields in two dimensional
Conformal Field Theory (CFT).   Whenever the OPE of some (primary)
fields is needed for dynamical analysis, the  associated CFT
provides the computational artillery and selects   the monomials
of the expansion.  Since a way of extending the bidimensional
conformal symmetry yields the notion of $W$-algebras, in this
context CFT arises as a perturbative groundstate around which one
expands the theory. $W$-algebras have been widely occurred in two
dimensional physics, such as the 2d anisotropic oscillator,
Coulombian and generalized hartmann potentials,
gravitation ($W$-gravity), condensed
matter (quantum Hall effect), integrable models (KdV, Toda), phase
transitions in two dimensions, intermediate statistics,
solitary waves since the intrinsic
intertwining between internal and space time symmetries realized
in these algebra   provide an amazing landscape to discuss
dynamics in a physical context.  The interested reader is referred
to some review papers such as \cite{Tjin,Capp,She92,Bou}, and many
others which have been implicitly quoted for the sake of brevity.

However, it turns out that the OPE mechanism is not the most natural
 way to get 
 an algebra. Indeed, the discovery of composition laws in products
between
quantum fields at coincident points, when inserted in Green functions
of complete set of states, do not allow to  
 define an algebra. Beyond them, further mathematical properties have to
be imposed, so that a lot of mathematical verifications, spread in the
literature, have to be performed. 
Anyhow the role played by CFT in all the  basic aspects of
$W$-algebras is crucial for these reasons and their mathematical
 aspects and  geometrical origin are important questions \cite{geom}.  

Furthermore, the distinctive feature of the (local) conformal symmetry
in two dimensions is 
that it provides an infinite number of conservation laws.
It turns out that this infinite symmetry has several consequences as was
first shown in \cite{BPZ}; a conformal symmetry is a (local)
diffeomorphism (i.e. a locally smooth and invertible map) which acts on
the
Hilbert space of the theory as a symmetry.

For all these above reasons one can ask whether a more canonical
approach, 
 such as the ones leading to Noether theorem in Lagrangian Field theory,
 or canonical transformations, may provide the construction of a
 certain type of $W$-algebras.
 Indeed, if we can find some kind of space-time transformations (due
 to the local  character of these algebras) where all the formal
requirements
 (associativity, existence of inverse and so on) 
 are fulfilled, then the usual calculation artillery of Field Theory
 allows a realization  of these algebras. In doing so, a larger insight
in the
 intrinsic nature of this symmetry could be achieved.    
  
Already Witten \cite{Wi1} has suggested that these algebras could have a 
space-time origin as symplectic diffeomorphisms, and Hull \cite{Hull1} 
gave some examples in terms of diffeomorphisms acting on fields (the
so-called $W$-matter).

In this paper we shall construct some of these $W$-algebras from 
{\em only coordinate symmetries} on the cotangent bundle.
In particular, we shall derive generating functions for a
very restrictive class of canonical transformations whose reduction 
will provide in general an infinite chain of smooth changes of
coordinates from a background and to new coordinate frames. 

In mathemathical terms, we shall study a realisation of the 
algebra of diffeomorphisms on the cotangent bundle over a world-sheet 
$\Sigma$, Diff$_0(T^{*}\Sigma)$, which extends the $W$-algebra.
A particular attention will be devoted to the complex structure mappings 
induced by these canonical transformations. 

In Section 2 we shall describe the geometrical approach which will lead 
to B.R.S transformations as they will be given in Section 3.

The general reduction to $W$-symmetry will be described in Section 4 
and in particular the $w_\infty$ symmetry will be derived.
A particular attention will be paid on the role of complex structure 
of each two dimensional graded space whose diffeomorphism algebra will 
generate the $W_n$ algebras. In particular the generalization of the
Beltrami 
differential parameters will also be discussed. Our approach will be 
complementary to those given in \cite{Zucchini,Gieres,Laz1}.

In order to illustrate the construction, we shall discuss in Section 5
the $W_3$ case. Two different diffeomorphism symmetries will give rise 
to two kinds of $W_3$-algebras respectively.
Firstly, a geometrical meaning will be given for the
$W_3$-algebra already discussed by Sorella et al \cite{Sorella} and
found through
the OPE mechanism by many authors \cite{many}. 
Secondly, it will be shown how the $W_3$ 
algebra given by Grimm et al \cite{Grimm}, and Ader and coworkers
\cite{Ader1}, \cite{Ader2} arises from the construction.

\sect{The Geometrical Approach}

\indent

Given a smooth compact 2d-surface $\Sigma$, we shall right away 
choose a prescribed local complex analytic coordinates $\zbz$. 
On the smooth cotangent bundle $T^*\Sigma$, we
will use throughout the paper local adapted complex coordinates
$(z,\bz;y_z,\ybz)$, with fibre coordinates $(y_z,\ybz)$
\cite{Kod86} for the natural coframe associated to the holonomic
coordinates $\zbz$. The former will be shorthandly denoted by $\zy$.
The canonical 1-form, $\theta$ on $T^*\Sigma$ then locally writes in
the local chart $\cU_\zy$ of $T^*\Sigma$,
\be
{\theta\arrowvert}_{\cU_\zy}\ =\ \yz dz  +\ybz d\bz .
\lbl{thetaz}
\ee
The fundamental 2-form, $\Omega\equiv d\theta$, reads in local
adapted coordinates on $T^*\Sigma$,
\be
{\Omega\arrowvert}_{\cU_\zy} 
\equiv d \yz \wedge dz  +  d \ybz \wedge d{\bz},
\lbl{omegaz}
\ee
and is closed. By Stokes theorem, one also has
\be
\int_{\Sigma} \Omega\ =\ \int_{\prt\Sigma}\theta, 
\ee
which vanishes if $\Sigma$ is without boundary.

Let us now consider a smooth change of local coordinates on $T^*\Sigma$,
$(Z,\bZ;\YZ,\YbZ)$, or $\ZY$ for short.
The canonical form then locally reads,
\be
{\theta\arrowvert}_{\cU_\ZY}\ =\ \YZ dZ + \YbZ d\bZ,
\lbl{thetay}
\ee
and the corresponding fundamental 2-form, 
\be
{\Omega\arrowvert}_{\cU_\ZY}\ =\ d\theta\ =\ d\YZ \wedge dZ  +
d\YbZ \wedge  d\bZ .
\lbl{omegaZ}
\ee
This smooth change of local coordinates on $T^*\Sigma$ turns out to
be a canonical transformation if the fundamental 2-form remains
invariant, 
which means that, on $\cU_\zy\cap \cU_\ZY$,
\be
{\Omega\arrowvert}_{\cU_\zy}={\Omega\arrowvert}_{\cU_\ZY},
\lbl{canonical}
\ee
This condition selects particular coordinate transformation laws 
which are usually called canonical transformations \cite{C-B}.
Eq.\rf{canonical} implies that on $\cU_\zy\cap \cU_\ZY$,
\be
{\theta\arrowvert}_{\cU_\zy}-{\theta\arrowvert}_{\cU_\ZY}\ =\ d F,
\ee
where $F$ is a generating function in the base coordinates
$(z,\bz,Z,\bZ)$ on 
$\pi\cU_\zy\cap \pi\cU_\ZY$ which is diffeomorphic to {\bf R}$^2\times$
{\bf
R}$^2$ and $\pi$ is the projection on $\Sigma$. 
It will be however more convenient to use the local coordinates 
$\zY$ by introducing the generating function $\Phi\zY$ through a
Legendre
transformation of $F$,
\be
d\Phi(z,Y)\equiv d \biggl (F(z,Z) + \YZ Z+\YbZ \bZ \biggr)
\ =\ \yz  dz + \ybz  d\bz + d\YZ Z + d\YbZ \bZ.
\ee
The function $\Phi(z,Y)$ is the generating function of the 
canonical transformation $\zy\lra\ZY$. It is locally defined (up to a
total
derivative) in the independent coordinates $\zY$ of the smooth
trivial bundle $\Sigma\times\mbox{\bf R}^2$ and has a non-singular
Hessian, ${\dps ||\frac{\prt^2 \Phi}{\prt z \prt Y}||\neq 0}$.
On $\Sigma\times\mbox{\bf R}^2$ the total differential is
\be
d=dz \prtz  +d \bz\prtbz   + d\YZ \frac{\prt}{\prt \YZ} +d\YbZ 
\frac{\prt}{\prt \YbZ}\equiv d_z + d_{\YZ},
\ee
and $d^2=0$ yields $d_z^2=d_{\YZ}^2=d_zd_{\YZ}+d_{\YZ}d_z=0$, or in
local
coordinates,
\be
\biggl[\prtz,\frac{\prt}{\prt \YZ} \biggr]=0,\qquad
\biggl[\prtbz,\frac{\prt}{\prt \YZ} \biggr]=0,
\lbl{commzy}
\ee
with the complex conjugate expressions.
Since $d^2\Phi=0$ we get the important identities\footnote{
From now on, we shall reserve $\prt$'s for $\prt\equiv\prtz$,
$\bprt\equiv\prtbz$.},
\be
\bprt \yz= \prt\ybz,\qquad
\frac{\prt}{\prt \YbZ} Z=\frac{\prt}{\prt \YZ} \bZ,\qquad
\frac{\prt}{\prt \YZ}\yz=\prt Z,\qquad
\frac{\prt}{\prt \YbZ}\yz=\prt \bZ, 
\lbl{integr2}
\ee
(and their c.c.). The generating function $\Phi$ induces the following
canonical transformation defined by,
\begin{eqnarray}
\zbz = \zbz,\qquad \yz\zY=\prt\Phi(z,Y), \lbl{Phi1}\\
Z\zY=\frac{\prt}{\prt \YZ} \Phi(z,Y),\qquad (Y_Z,\YbZ)=(Y_Z,\YbZ),
\lbl{Phi2}
\end{eqnarray}
with also the complex conjugate coordinates. 
In account of this choice of generating function, $(\zbz;\yz,\ybz)$
becomes a
family of local smooth sections of the cotangent bundle $T^*\Sigma$
parametrized by $Y$ and over $\pi\cU_\zy\cap \pi\cU_\ZY$.

As is well known, by the implicit function theorem, eq.\rf{Phi1} is 
locally solved in $Y\zy$ and plugging into Eq.\rf{Phi2} one ends with 
\be
Z\zy\ =\ \left(\frac{\prt}{\prt \YZ}
\Phi(z,Y)\right)\arrowvert_{Y=Y\zy}.
\lbl{Ztrue}
\ee
One also checks that
\be
{\theta\arrowvert}_{\cU_\zy}\ =\ d_z \Phi(z,Y)\arrowvert_{Y=Y\zy},
\lbl{Phi}
\ee
where the 1-form, $d_z \Phi(z,Y)$, on $\Sigma\times\mbox{\bf R}^2$,
is evaluated on a solution $Y\zy$ of eq.\rf{Phi1}.
Accordingly, the fundamental 2-form simply reads,
\be
{\Omega\arrowvert}_{\cU_\zy} \ =\ d{\theta\arrowvert}_{\cU_\zy}\ =\
d_y\biggl(d_z \Phi(z,Y)\arrowvert_{Y=Y\zy}\biggr)
\ =\ \biggl(d_Y d_z\Phi\zY\biggr)\arrowvert_{Y=Y\zy},
\lbl{omegafull}
\ee
where the r.h.s. may be viewed as the restriction onto solutions $Y\zy$
of 
a 2-form exact on each factor of the product $\Sigma\times\mbox{\bf
R}^2$.
An important remark is in order. Due to $d^2\Phi=0$, 
the 2-forms $d_{\YZ} Z \wedge d{\YZ} + d_{\YZ}\bZ\wedge d{\YbZ}$ 
or $d_{z}\yz \wedge dz + d_z \ybz \wedge d\bz $ identically vanish in 
$\Omega$. This yields two very particular classes of
canonical transformations, in particular, those which allow
reparametrizations of $Z\zY$ in the $\YZ$ fibre coordinate only. One has
the
following theorem.

\newtheorem{guess}{Theorem}
\begin{guess} \label{thm}
On the smooth trivial bundle $\Sigma\times\mbox{\bf R}^2$,
the vertical holomorphic change of local coordinates, 
\begin{eqnarray}
Z(\zbz ,\YbY) \lra Z(\zbz, \cF(Y_Z),\YbZ),
\end{eqnarray}
where $\cF$ is a holomorphic function in $\YZ$, while the horizontal
holomorphic change of local coordinates,
\begin{eqnarray} 
\yz(z,\bz,\YbY)\lra \yz(f(z),\bz,\YbY),
\end{eqnarray}
where $f$ is a holomorphic function in $z$,
are both canonical transformations.
\end{guess}
The latter type of canonical transformations states that the
fundamental 2-form remains unchanged under local holomorphic changes of
the
local $z$-coordinate on the basis, namely holomorphic changes of charts. 
In the former, one can restrict oneself to a very particular
situation. Since any smooth change of local complex coordinates on 
the base Riemann surface
$\Sigma$, $\zbz\lra (Z\zbz,\bZ\zbz)$, can be obtained by the generating
function,
\be
\Phi\zY\ =\ Z\zbz Y_Z + \bZ\zbz\YbZ,
\lbl{zZ}
\ee
one may consider more generally on $\Sigma\times\mbox{\bf R}^2$ 
the following holomorphically split
generating function in the vertical direction, namely,
\be
\Phi\zY\ =\ \Phi_1(\zbz,\YZ) +\ovl{\Phi}_1(\zbz,\YbZ).
\ee

Considering the expansion around $\YZ=\YbZ=0$, the 
generating function $\Phi$ will be written in formal power series, 
\begin{eqnarray}
\Phi\zY=\sum_{n\geq 1}
\biggl[\frac{1}{n!}\,\YZ^n{\biggl(\frac{\prt}{\prt \YZ}\biggr)}^n
\Phi_1(z,Y)\arrowvert_{\YZ=0}\biggr]
+ \sum_{n\geq 1} \biggl[\frac{1}{n!}\,\YbZ^n {\biggl(\frac{\prt}{\prt
\YbZ}\biggr)}^n\ovl{\Phi}_1(z,Y)\arrowvert_{\YbZ=0}\biggr]\nn \\[2mm]
\equiv
\sum_{n\geq 1}\biggl[\YZ^n\, \Zn\zbz\biggr]+\sum_{n\geq 1}\biggl
[\YbZ^n\, \bZ^{(n)}\zbz\biggr]
\lbl{phi}
\end{eqnarray} 
where we have set $\Phi\zY\left|_{\YZ,\YbZ=0}\right.=0$\footnote{Setting 
rather $\Phi\zY\left|_{\YZ,\YbZ=0}\right.= \Phi_0\zbz$, induces that
1-form $y$ is defined up to the coboundary $d\Phi_0$. Also $\Phi$ is
connected to the identity map $z\YZ + \bz\YbZ$.}, and introduced
together with its complex conjugate counterpart,
\be
\Zn\zbz\equiv\biggl[\frac{1}{n!}{\biggl(\frac{\prt}{\prt \YZ}\biggr)}^n
\Phi_1\zY\biggr]
\arrowvert_{\YZ=0}
\lbl{Zn}
\ee
as the jet coordinates of $\Phi$ at $Y=0$ with respect to $Y$
parametrized by $\zbz$. Making use of these jet coordinates will provide
a new scenario for generating $W$-algebras.
Note also that taking $Y=0$ implies $y=0$.

We shall now introduce some quantities which will be useful in the
sequel.
Let us denote,
\begin{eqnarray}
\lambda\zY =\prtz\prtY\Phi\zY,\qquad
\lambda\zY\mu\zY=\prtbz\prtY\Phi\zY,\nn\\
\blambda\zY=\prtbz\prtbY\Phi\zY,\qquad
\blambda\zY\mub\zY=\prtz\prtbY\Phi\zY.
\lbl{parameters}
\end{eqnarray}
Before being restricted to the solutions $Y=Y\zy$, the globally
defined fundamental 2-form \rf{omegafull} rewrites in the $\zY$ 
independent local coordinates,
\begin{eqnarray}
\Omega\zY&=& \biggl(\lambda 
d\YZ+\blambda\mub d\YbZ\biggr)\wedge dz +
\biggl(\blambda d\YbZ+\lambda\mu d\YZ\biggr) \wedge d\bz \nn\\
&=& d\YZ\wedge\lambda\biggl(dz +\mu d\bz\biggr) 
+ d\YbZ\wedge \blambda \biggl( d \bz + \mub dz \biggr)
\end{eqnarray}
from which one can read off by construction,
\begin{eqnarray}
d_z Z\zY&=&\lambda\zY\biggl(dz +\mu\zY d\bz\biggr)\nn\\[-2mm]
\lbl{dd}\\
d_Y\yz\zY&=&\lambda\zY\biggl(
d\YZ+\frac{\blambda\zY\mub\zY}{\lambda\zY} d\YbZ\biggr)\nn
\end{eqnarray}
(and the c.c. expressions).
The latter reveal a parametrization of complex  structures by 
a Beltrami  differential depending on the vertical
coordinate $Y$, $\mu\zY$, on the base surface $\Sigma$ according to
the complex coordinates $\zbz$, and
by ${\dps \frac{\blambda\zY\mub\zY}{\lambda\zY}}$ on the vertical fiber
when $\YbY$ complex coordinates are used.
The restriction condition $ d\Omega\arrowvert_{Y=Y\zy}=0 $ gives,
\begin{eqnarray}
d {\Omega\arrowvert}_{\cU_\zY} = \bprt \lambda\zY d\bz
\wedge dz\wedge d\YZ +\bprt(\blambda\zY \mub\zY) d\bz\wedge dz\wedge
d\YbZ\nn \\ 
+\ \prt(\lambda\zY\mu\zY)dz\wedge d\bz\wedge d\YZ + \prt \blambda\zY 
dz\wedge d\bz\wedge
d\YbZ \nn \\
 +\ d\YbZ\wedge \frac{\prt}{\prt \YbZ}
\biggl( \lambda\zY\biggl(dz +\mu\zY d\bz\biggr)\biggr)\wedge d\YZ\nn \\
+\ d\YZ\wedge \frac{\prt}{\prt \YZ}
\biggl( \blambda\zY\biggl(d\bz +\mub\zY dz\biggr)\biggr)\wedge d\YbZ=0,
\end{eqnarray}
and yields the Beltrami identities,
\begin{eqnarray}
\bprt \lambda\zY =\prt(\mu\zY \lambda\zY), \qquad
\frac{\prt}{\prt \YbZ} \lambda\zY = \frac{\prt}{\prt \YZ} \biggl
(\blambda\zY \mub\zY\biggr)
\lbl{beltra}
\end{eqnarray}
together with their complex conjugate expressions.
The Liouville theorem follows from,
\begin{eqnarray}
\det \left|\frac{\prt Z\zY}{\prt z}\right|
=\lambda\zY\blambda\zY(1-\mu\zY\mub\zY)
 = \det \left|\frac{\prt y\zY}{\prt Y}\right|.
\end{eqnarray}
Therefore, one can consider two distinct smooth changes of local
coordinates which are indeed 
strictly related through the given generating function
$\Phi$. The former on the base,
$\zbz\longrightarrow\ZBZ$ for fixed $(\YZ,\YbZ)$, 
and the latter on the fiber over $\zbz$,
$(\yz,\ybz)\longrightarrow (\YZ,\YbZ)$,

Accordingly, the transformation laws of the derivative operators on
each factor of
the base or on the cotangent fiber respectively read,
\begin{eqnarray}
\prtZ &=& \frac{\prtz-\mub\zY\prtbz}{\lambda\zY(1-\mu\zY\mub\zY)},
\\[2mm]
\frac{\prt}{\prt\YZ}\arrowvert_{Y=Y\zy}
&=&\lambda\zY\biggl(\frac{\prt}{\prt\yz}+
\mu\zY\frac{\prt}{\prt\ybz}\biggr)\arrowvert_{Y=Y\zy} \equiv
\lambda\zY\arrowvert_{Y=Y\zy} \cD^z,
\lbl{prt}
\end{eqnarray}
(with their c.c. expressions) and where the combination $\cD^z$ of
the vertical derivatives with respect to the $(\yz,\ybz)$ fiber 
coordinates has been introduced. First, it has to be noted that
Eqs.\rf{beltra}
infer the following important identity,
\begin{eqnarray}
\cD^z \ln\blambda\zY\arrowvert_{Y=Y\zy} 
=
\frac{\cbD^\bz \mu\zY\arrowvert_{Y=Y\zy}
+\mu\zY\arrowvert_{Y=Y\zy}\cD^z\mub\arrowvert_{Y=Y\zy}\zY}{1-\mu\zY\arrowvert_{Y=Y\zy}
\mub\zY\arrowvert_{Y=Y\zy}},
\lbl{beltraD}
\end{eqnarray}
with of course the c.c. formula, and secondly, from Eq.\rf{commzy} one
gets,
\begin{eqnarray}
\biggl[\prtz,\cD^z\biggr]&=&
-\prtz\log\lambda\zY\arrowvert_{Y=Y\zy}\cD^z,\nn\\
\biggl[\prtbz,\cD^z\biggr]&=&
-\prtbz\log\lambda\zY\arrowvert_{Y=Y\zy}\cD^z\\
&=&-\biggl(\prt\mu\zY\arrowvert_{Y=Y\zy}\cD^z-\mu\zY\arrowvert_{Y=Y\zy}
\biggl[\prtz,\cD^z\biggr]\biggr),\nn
\end{eqnarray}
(and the c.c. expressions), and,
\be
\biggl[\cD^z,\cbD^\bz\biggr]\ =\ 
(\cD^z\mub\zY\arrowvert_{Y=Y\zy})\frac{\prt}{\prt\yz}-(\cbD^\bz\mu\zY
\arrowvert_{Y=Y\zy})\frac{\prt}{\prt\ybz},
\ee
from which one verifies that $\yz$ and $\ybz$ turn out to be
independent fiber coordinates, 
\be
\biggl[\frac{\prt}{\prt\yz},\frac{\prt}{\prt\ybz}\biggr]\ =\ 0.
\ee

\sect{The smooth diffeomorphism action}

\indent

So far we have considered canonical transformations as
changes of local coordinates on the cotangent bundle $T^*\Sigma$ while
those generated by holomorphically split generating functions in the
vertical coordinates have been preferred. Since our strategy amounts
to choosing $\zY$ as independent local coordinates 
with generating function $\Phi\zY$, it is recalled that 
$y_z=\prt\Phi\zY$ defines a section of the cotangent bundle $T^*\Sigma
\stackrel{\pi}{\lra} \Sigma$. One may wonder how $\Phi\zY$ is
 affected by the action of smooth diffeomorphisms of the
trivial bundle $\Sigma\!\times\!\mbox{\bf R}^2$ 
as a manifold with coordinates $\zY$. 
In order to implement the diffeomorphism
action, the BRS differential algebra setting will be used, see
e.g. \cite{Ba88}. 
We consider the action of smooth diffeomorphisms homotopic to the
identity map, $\varphi_t = id_{\Sigma} + t\, c + o(t)$, where 
$c = \xi^z\zY\prt + \xi^{\ovl{z}}\zY\ovl{\prt} 
+ \eta_Z\zY\frac{\prt}{\prt\YZ} +
\ovl{\eta}_{\ovl{Z}}\zY\frac{\prt}{\prt\YbZ} \equiv \xi\!\cdot\!\prt +
\eta\!\cdot\!\frac{\prt}{\prt Y}$, is the smooth Faddeev-Popov
ghost associated to vector fields\footnote{Mathematically speaking, 
$c$ is the generator of the Grassmann algebra of the dual of the Lie 
algebra of smooth diffeomorphisms.}
with respect to the background complex coordinates $\zY$, $s\,z=s\,Y=0$. 
The corresponding infinitesimal action on fields over the cotangent
bundle is obtained by a (graded) Lie derivative encoded in a 
nilpotent BRS operation $\sS\equiv L_c = i_c\, d - d\, i_c$, $\sS^2=0$. 

Then at the infinitesimal level, the BRS variation of $\Phi\zY$
locally writes,  
\be
\sS\Phi\zY\ =\ L_c \Phi\zY\ =\  
\xi\zY\cdot y\zY + \eta\zY\cdot Z\zY \equiv \Lambda\zY,
\lbl{lambda}
\ee
where $\Lambda\zY$ is a Grassmann function subject to $\sS
\Lambda\zY=0$.
The variation of the fundamental 2-form \rf{omegafull} 
in $\cU_{\zy}\cap \cU_{\ZY}$ writes,
\be
\sS\Omega\zY\ =\ d_{\YZ} d_z\Lambda\zY,
\ee
and for the canonical 1-form one has, in virtue of Eq.\rf{Phi},
\begin{eqnarray}
\sS{\theta\arrowvert}_{\cU_{\zy}}&=&d_z\Lambda\zY\arrowvert_{Y=Y\zy}\nn
\\
\sS{\theta\arrowvert}_{\cU_{\ZY}}&=&\biggl[d\biggl(\YZ\frac{\prt}{\prt
\YZ}\Lambda\zY+ \YbZ \frac{\prt}{\prt \YbZ}\Lambda\zY\biggr)
-d_{\YZ}\Lambda\zY\biggr]\arrowvert_{z=z(Z,Y)}
\lbl{theta1}
\end{eqnarray}
in such a way that the invariance of $\Omega$ and $\theta$ writes,
\be
\sS\int_{\Sigma} \Omega\ =\ \sS\int_{\prt\Sigma}\theta\ =\ 0.
\ee
To the general canonical transformations given by Eqs.\rf{Phi1} and
\rf{Phi2} preserving the symplectic form $\Omega$
as stated in Eq.\rf{canonical}, there will correspond the
infinitesimal variations respectively, 
\bea
\sS\yz\zY=\prtz\Lambda\zY, \qquad \qquad
\sS Z\zY=\frac{\prt}{\prt \YZ}\Lambda\zY
\equiv\Upsilon^Z\zY,
\lbl{ZZ}
\ena
and their complex conjugates. 
The latter is the infinitesimal transformation of the new local
holomorphic coordinate $Z$ in terms of the ghost vector 
$c = \Upsilon^Z \prtZ + \Upsilon^{\ovl{Z}} \prt_{\ovl{Z}}$ expressed in
the new system $\ZBZ$. We also have,
\begin{eqnarray}
\sS \lambda\zY=\frac{\prt}{\prt \YZ}\prtz\Lambda\zY, \qquad
\sS \biggl(\blambda\zY \mub\zY\biggr) =
 \frac{\prt}{\prt \YbZ}\prtz\Lambda\zY. 
\end{eqnarray}
In order to restore the explicit dependence in the $Y$ coordinates,
from Eqs.\rf{ZZ} and \rf{prt}, one defines,
\begin{eqnarray}
\sS Z\zY\equiv\Upsilon^Z\zY =
\frac{\prt}{\prt \YZ}\Lambda\zY
\equiv \lambda\zY \cC\zY ,
\end{eqnarray}
so that,
\begin{eqnarray}
\cC\zY= \frac{1}{\lambda\zY}\Upsilon^Z\zY
\lbl{cC1}
\end{eqnarray}

We now restrict ourselves to the solutions $Y=Y\zy$ according to the
strategy 
defined in Eq.\rf{prt},
\begin{eqnarray}
\cC\zY\arrowvert_{Y=Y\zy}&=&\cD^z\biggl(\Lambda\zY\arrowvert_{Y=Y\zy}\biggr)
\nn \\[2mm]
&=& c\zy +\mu\zY\arrowvert_{Y=Y\zy} \bc\zy,
\lbl{ghosts}
\end{eqnarray}
where we have set,
\be
c\zy=\frac{\prt}{\prt \yz}\biggl(\Lambda\zY\arrowvert_{Y=Y\zy}\biggr),
\qquad
\bc\zy=\frac{\prt}{\prt
\ybz}\biggl(\Lambda\zY\arrowvert_{Y=Y\zy}\biggr).
\lbl{ghosts1}
\ee
Thus Eq.\rf{cC1} when restricted to the solutions $Y=Y\zy$
is given by \rf{ghosts} in terms of a derivative of 
$\Lambda$ with respect the background $(\yz,\ybz)$ local fiber
coordinates. This ansatz is of course strongly supported by 
the implicit function theorem which is supposed to be understood from
now. 
Moreover, one has the following identity,
\begin{eqnarray}
\biggl[\sS,\cD^z\biggr]&=&-\cC\zY\prtz\log\lambda\zY\cD^z-\prt
\cC\zY\cD^z\nn\\
&=&\cC\zY\biggl[\prt,\cD^z\biggr]-\prt \cC\zY\cD^z
\end{eqnarray}
which yields the variations,
\begin{eqnarray}
\sS\lambda\zY&=&\prt\biggl(\lambda\zY \cC\zY\biggr)\\
\lbl{slambda1}
\sS \mu\zY&=&\cC\zY \prt\mu\zY -\mu\zY\prt \cC\zY +\bprt \cC\zY \\
\lbl{smu}
\sS \cC\zY&=& \cC\zY\prt \cC\zY \\
\lbl{sC}
\sS c\zy&=&(c\zy\prt+\bc\zy\bprt)c\zy  
\lbl{sc}
\end{eqnarray}
(and their c.c. expressions).

\sect{Towards a $W$-algebra presentation}

\indent

We have previously chosen canonical transformations 
described by the change of local complex coordinates parametrized by
$Y$,
\begin{eqnarray}
Z\zbz\lra  Z(\zbz,\YZ) = Z\zbz + \sum_{n\geq 2} n\,\Zn \zbz{\YZ}^{n-1}
\lbl{Zseries}
\end{eqnarray}
where $\Zn\zbz$ has been defined in Eq.\rf{Zn}.
The choice of $\YZ$ as an independent variable,
accordingly gives that the mappings,
\begin{eqnarray}
\zbz \lra \ZBZn,\qquad n\geq 1,
\end{eqnarray}
generate a tower of smooth changes of local complex 
coordinates on the base Riemann
surface, each of them will be shown to be local solutions of a Beltrami
like equation.

\subsect{The complex structures underlying the $\Zn$ local complex
coordinates}

\indent

It is now shown how each $\ZBZn$ coordinates define new complex
coordinates pertaining to a complex structure. 	Indeed, by construction
one can write,
\begin{eqnarray}
d_z \Zn\zbz &=& \biggl
[\frac{1}{n!} {\biggl(\frac{\prt}{\prt \YZ}\biggr)}^n\prt\Phi\zY
dz +\frac{1}{n!} {\biggl(\frac{\prt}{\prt
\YZ}\biggr)}^n\bprt\Phi\zY\biggr]
\left|_{\YZ=0,\YbZ=0}\right.\nn \\
&\equiv& \lambdan\zbz [dz +\mu_\bz^z(\zbz,n) d\bz] 
\end{eqnarray}
where we have introduced
\begin{eqnarray}
\lambdan\zbz\equiv \biggl[\frac{1}{n!} {\biggl(\frac{\prt}{\prt \YZ}
\biggr)}^n\prt\Phi\zY\biggr]\arrowvert_{\YZ=0,\YbZ=0}\equiv \prt \Zn\zbz
\\
\lambdan\zbz \mu_\bz^z(\zbz,n) \equiv
\biggl[\frac{1}{n!} {\biggl(\frac{\prt}{\prt
\YZ}\biggr)}^n\bprt\Phi\zY\biggr]
\arrowvert_{\YZ=0,\YbZ=0}
\equiv \bprt \Zn\zbz.
\lbl{lambdan}
\end{eqnarray}
Since the generating function is supposed to be complex analytic in
$Y$, both the convergence of the  series \rf{phi} and the  requirement
that the mappings, $\zbz \lra \ZBZn$, preserve the orientation
lead to the condition $|\mu_\bz^z(\zbz,n)|\le 1$.

A Beltrami identity is immediately recovered for each level $n$,
\begin{eqnarray}
\biggl[\frac{1}{n!} {\biggl(\frac{\prt}{\prt
\YZ}\biggr)}^n\prt\bprt\Phi\zY
\biggr]
\arrowvert_{\YZ=0,\YbZ=0}=
\bprt\lambdan\zbz= \prt(\lambdan\zbz\mu_\bz^z(\zbz,n))
\end{eqnarray}
The expression for $\lambdan$ is  non local in $\mu_\bz^z(\zbz,n)$.
It is now clear that the quantity $\mu_\bz^z(\zbz,n)$ encodes
the complex structure of the space $\Zn$. From the very definitions, a
direct computation yields the following expansion
\begin{eqnarray}
\mu_\bz^z(n,\zbz) 
\ =\ \sum_{k=1}^{n} \omega_{(k-1)}(n,\zbz)\,\mu^{(k)}_\bz\zbz,
\lbl{mun}
\end{eqnarray}
in terms of local $(-n,1)$-conformal fields,
($\mu_\bz^z(n,\zbz)=\mu^{(1)}_\bz\zbz \equiv \mu\zbz$ for $n=1$),
\begin{eqnarray}
\mu_{\bz}^{(n)}\zbz=\biggl[
\frac{1}{n!}{\biggl(\cD^z\biggr)}^{n}\bprt\Phi\zY\biggr]
\left|_{\YZ,\YbZ=0}\right.
\equiv\biggr[
\frac{1}{n!}{\biggl(\cD^z\biggr)}^{n-1}\mu\zY\biggr]\left|_{\YZ,\YbZ=0}\right.
\lbl{mun0}
\end{eqnarray}
and with coefficients $(k-1,0)$-conformal fields, non local in the
$\mu_\bz^z(\zbz,m)$ of orders $m\leq n$,
\begin{eqnarray}
\omega_{(k-1)}(n,\zbz) \equiv k! \left.\left(
\prod_{j=1}^k
\frac{{\lambdapj}^{a_j}\zbz}{a_j!\,\lambdan\zbz}\right)
\right|_{\tiny {\dps \begin{array}{c} {\dps \sum_{i=1}^k a_i =k,
\mbox{ and } \sum_{i=1}^k p_i\,a_i = n,}\\ 
\mbox{with } n\geq p_1>\cdots>p_n\geq 0. \end{array}}}
\lbl{omegak}
\end{eqnarray}
with $\omega_{0}\zbz = 1$. From Eq.\rf{mun} we can state the Theorem,
\begin{guess}
The complex structure of the $\ZBZn$  can  be 
 described   by 
parameters     $\mu_\bz^z(\zbz,n)$    
which extend to these spaces the  
Beltrami multipliers. The parameters $\mu_\bz^z(\zbz,n)$ (for a given $n$), 
depend in a  non local way on their partners $\mu_\bz^z(\zbz,j)$
for $j \leq n$.
\end{guess}

This important geometrical statement  
it is the basis for the 
physical discussion of the problem.

The previous arguments show that the quantities $\mu_{\bz}^{(n)}\zbz$ 
parametrize the change of the complex structure of the $\ZBZn$ 
in terms of the index $n$.

The role of the parameters $\mu_\bz^{(j)}(\zbz)$ can be understood 
from Eq \rf{mun} since it is easy to realize that the  coordinate
system $\ZBZn$ will have a local complex structure $\mu$ as  
$\ZBZ$ does, if and only if $\mu_\bz^{(j)}(\zbz)=0$ for $ 2\leq j<n$.

Obviously few examples can better clarify this point; this will be
performed 
in the most simple cases in the next Section.
 
We do not pretend to exhaust and to classify all the $W$ algebras in our 
context (more examples and a deeper insight in  
critical situations will be needed), but we hope to reduce to an unique 
geometrical setting the more common $W$ algebras studied in the
literature.   

\subsect{The $W$-symmetry}

\indent

We derive now a $W$-symmetry from the previous construction combining
both the diffeomorphism action and the canonical transformations via the 
B.R.S machinery. For each $n$ we define the diffeomorphism action on
the local complex coordinate $\Zn$ by,
\begin{eqnarray}
\sS \Zn\zbz\equiv\Upsilon^{(n)}\zbz\equiv\biggl[\frac{1}{n!}
{\biggl(\frac{\prt}{\prt
\YZ}\biggr)}^n\Lambda\zY\biggr]\left|_{\YZ,\YbZ=0}\right.
\end{eqnarray}
which are invariant ghost functions since the $\sS$ nilpotency  gives:
\begin{eqnarray}
\sS\Upsilon^{(n)}\zbz=0.
\end{eqnarray}
In more details, one has
\begin{eqnarray}
\Upsilon^{(n)}\zbz &=& \biggr[
\frac{1}{n!} {\biggl(\frac{\prt}{\prt \YZ}\biggr)}^{n-1}
(\lambda\zY \cC\zY)\biggr]\left|_{\YZ,\YbZ=0}\right.\nn\\
&=& \lambdan\zbz \sum_{k=1}^{n} \omega_{(k-1)}(n,\zbz)\, \cC^{(k)}\zbz
\end{eqnarray}
where we have introduced the following $(-n,0)$-conformal ghost fields,
\begin{eqnarray}
\cC^{(n)}\zbz 
=\biggl[\frac{1}{n!} {\biggl(\cD^z\biggr)}^n \Lambda\zY\biggr]
\left|_{\YZ,\YbZ=0}\right., \qquad n=1,2,\cdots, \qquad
\cC^{(1)}\zbz\equiv \cC \zbz,
\lbl{0000000b}
\end{eqnarray}
and where the non local coefficients $\omega_{(k-1)}(n,\zbz)$ have
been previously introduced in Eq.\rf{omegak}.

 From the very definitions one obtains the obvious law of 
transformation under diffeomorphisms,
\begin{eqnarray}
\sS \cC^{(n)}=\sum_{r=1}^n r\,\cC^{(r)}\prtz \cC^{(n-r+1)}.
\lbl{000b}
\end{eqnarray}
Moreover, one can see that the $(-n,1)$-conformal fields \rf {mun0}
are also given by,  
\be
\mu^{(n)}_\bz\ =\ \frac{\prt}{\prt\bc} \cC^{(n)},
\ee
so that by a trick related to diffeomorphisms
\cite{BaLa93}, namely,
${\dps \left\{\cS,\frac{\prt}{\prt\bc}\right\} = \bprt,}$
one easily obtains their BRS variations,
\be
\sS \mu_{\bz}^{(n)}\zbz=\bprt C^{(n)}\zbz + \sum_{r=1}^n 
r \biggl(C^{(r)}\zbz \prt\mu_{\bz}^{(n-r+1)}\zbz -
\mu_{\bz}^{(r)}\zbz \prt C^{(n-r+1)}\zbz \biggr).
\ee
For the non local fields one gets,
\begin{eqnarray}
\sS\lambdan\zbz&=&\biggl[\frac{1}{n!} {\biggl(\frac{\prt}{\prt \YZ}
\biggr)}^n\prt\Lambda\zY\biggr]\left|_{\YZ,\YbZ=0}\right.\nn\\
&=&\biggl[\prt\biggl(
\frac{1}{n!} {\biggl(\frac{\prt}{\prt \YZ}
\biggr)}^{(n-1)}(\lambda\zY\cC\zY)\biggr)\biggr]\left|_{\YZ,\YbZ=0}\right.
\nn \\ 
&\equiv& \prt\biggl(\lambdan\zbz \cK^z(\zbz,n)\biggr)
\end{eqnarray}
where the new ghost vector field,
\begin{eqnarray}
\cK^z(\zbz,n)\ =\ \sum_{k=1}^{n} \omega_{(k-1)}(n,\zbz)\,\cC^{(k)}\zbz
\lbl{kn1}
\end{eqnarray}
has as variation,
\begin{eqnarray}
\sS\cK^z(\zbz,n)=\cK^z(\zbz,n)\prt\cK^z(\zbz,n),
\lbl{skn}
\end{eqnarray}
and since by construction,
\begin{eqnarray}
\sS\biggl[ \lambdan\zbz \mu_\bz^z(\zbz,n)\biggr] = \sS \bprt Z^{(n)}
=\bprt \Upsilon^{(n)} = \bprt (\lambdan\zbz\cK^z(\zbz,n)),
\end{eqnarray}
one finally gets,
\begin{eqnarray}
\sS  \mu_\bz^z(\zbz,n) = \cK^z(\zbz,n)\prt\mu_\bz^z(\zbz,n)
-\mu_\bz^z(\zbz,n)\prt
\cK^z(\zbz,n)+\bprt\cK^z(\zbz,n).
\end{eqnarray}
For each $n$ a diffeomorphism BRS-structure with a ghost $\cK^z(\zbz,n)$ 
(non local in the complex structure parameters) can be put into
evidence.

While  the case  $n=1$ 
identifies the diffeomorphism symmetry, we show now that 
for each sector of the grading 
$n=1\cdots$ we shall individuate a $W$ symmetry.
 
The new ghost fields defined by:
\begin{eqnarray}
c^{(p,q)}\zbz 
=\biggl[\frac{1}{p!}\frac{1}{q!} {\biggl(\frac{\prt}{\prt \yz
}\biggr)}^p {\biggl(\frac{\prt}{\prt \ybz}\biggr)}^q
\Lambda\zY\biggr]\left|_{\YZ,\YbZ=0}\right.
\lbl{0000000a}
\end{eqnarray}
with $c^{(1,0)}\zbz=c^z(z,y)|_{y=0}$, transform as,
\begin{eqnarray}
\sS c^{(p,q)}\zbz&=&\sum_{\tiny {\dps \begin{array}{l} r=0,\dots,p\\
s=0,\dots,q\\ r+s>0\end{array}}}
\biggl(r\,c^{(r,s)}\zbz\prtz
c^{(p-r+1,q-s)}\zbz  \nn \\
&&\qquad\qquad +\ s \,c^{(r,s)}\zbz\prtbz
c^{(p-r,q-s+1)}\zbz\biggr).
\lbl{0000a}
\end{eqnarray}
Notice that the variations  of $c^{(p,q)}$  contain 
the fields $c^{(r,s)}$ with degrees $r\leq p$ and $s\leq q$.
The previous transformations \rf{0000a} induce the following
commutation relations, for currents defined through the nilpotent
functional BRS operator, 
\be
\delta \equiv
\int_{\Sigma} d\bz\wedge dz \biggl(c^{(p,q)}\zbz \cT_{(p,q)}\zbz + \sS
c^{(p,q)}\zbz \frac{\delta}{\delta c^{(p,q)}} \biggr),
\ee
namely,
$\{\delta,\delta\} = 0$, leads to,
\begin{eqnarray}
\biggl[\cT_{(p,q)}\zbz, \cT_{(r,s)}\zbzp\biggr]
&\!\!=\!\!& p\, \prt_{z'}\delta^{(2)}(z'-z)\cT_{(p+r-1,q+s)}\zbz 
- r\, \prtz\delta^{(2)}(z-z') \cT_{(p+r-1,q+s)}\zbzp \nn\\[2mm]
&& \hskip -3.5cm
+\ q\,\bprt_{z'} \delta^{(2)}(z'-z) \cT_{(p+r,q+s-1)}\zbz
- s\, \prtbz \delta^{(2)}(z-z') \cT_{(p+r,q+s-1)}\zbzp.
\lbl{02}
\end{eqnarray}
which turn out to be a realization of the so-called
$W_{\infty}$-algebra according to 
\cite{Hull1}, if no limit is put on the grading indices. Otherwise, if
a truncation criterium is given (by fixing a suitable upper limit 
on the series in Eq.\rf{phi}) a $W_n$-algebra (with $n$ obviously 
finite) can be constructed as will be seen  in the next section for
the most simple examples.

\sect{The $W_3$ example}

\indent

In the literature there exist two types of the named $W_3$-algebra,
namely the former is obtained by a field realization
\cite{Hull1,Sorella} which is on-shell, while the latter is given by
reduction
\cite{geom,many,OSSN,Grimm}. It will be shown down below how those two
realizations of $W_3$-algebra, will be re-obtained from our
construction only grounded on the combination of canonical
transformations and diffeomorphisms. 

\subsect{The chiral $W_3$-gravity}

\indent

We have seen in the preceding section how the sequences of 
smooth changes of local complex coordinates,
\begin{eqnarray}
\zbz\lra \ZBZn, \qquad
n=1,2\cdots \infty
\end{eqnarray}
generate the $W_\infty$ algebra. By the way we can arbitrarily
truncate the series to a non empty  
subset (or at least only one) of these invariance laws and see the 
physical consequences it will imply. The  simplest example
is obtained by considering only the smooth change of local complex
coordinates, 
\begin{eqnarray}
\zbz \lra \ZBZdue,
\end{eqnarray}
while neglecting the previous one of the lower order,
\begin{eqnarray}
\zbz \lra \ZBZ.
\end{eqnarray}
This will show, upon imposing the diffeomorphism invariance,
the coordinate origin of the chiral $w_3$ algebra
previously obtained by Sorella et al \cite{Sorella}.

Following the notation of the previous Section we have successively,
\begin{eqnarray}
\lambdadue=\prt Z^{(2)},\qquad
\lambda=\prt Z,\\
\sS Z^{(2)}=\Upsilon^{(2)}=\lambdadue \cC+\lambda^2\cC^{(2)}\equiv 
\lambdadue\cK^{z}(2),\nn \\[2mm]
\sS\lambdadue=\prt\Upsilon^{(2)}=\prt\biggl(\lambdadue\cK^{z}(2)\biggr)\\
\cK^{z}(2)=\cC +\frac{{(\lambdaZ) }^2}{\lambdadue} \cC^{(2)}\equiv
\cC +\frac{\beta_z}{2} \cC^{(2)},
\lbl{cK}
\end{eqnarray}
where we have set 
 \begin{eqnarray}
\beta_z=\frac{2{(\lambdaZ)}^2}{\lambdadue}.
\end{eqnarray}
The nilpotency condition on $\sS Z^{(2)}$ gives,
\begin{eqnarray}
\sS \cK^{z}(2)=\cK^{z}(2)\prt\cK^{z}(2).
\lbl{sK}
\end{eqnarray}
The decomposition of the ghost $\cK^{z}(2)$ turn out to be
 so useful to see how the underlying coordinate $Z$ behaves under the
 cotangent diffeomorphisms. 

In fact while the $\cC$ term is the usual ghost of the $\zbz\lra\ZBZ$
mapping,
the $\frac{\beta_z}{2} \cC^{(2)}$ term  upgrades the diffeomorphism to
the 
$\zbz\lra\ZBZdue$ level.
The parameter ${\beta_z}$ does contain the relative change of the $Z$ 
coordinate with respect  to the   background $\zbz$
used to describe the $\ZBZdue$ invariance.
So the independent study of the  behaviour of $\cK^z(2)$ and 
$\frac{\beta_z}{2} \cC^{(2)}$ under   Eq \rf{sK}  gives more details
on the realization of  the diffeomorphism.
In order to have a more precise geometrical information we shall fix
the $\cC^{(2)}$ BRS transformation to be,
\begin{eqnarray}
\sS \cC^{(2)} = \cC\prt\cC^{(2)} +2\cC^{(2)}\prt\cC,
\lbl{sc2}
\end{eqnarray}
in order to deduce the consistent transformation of $\beta_z$ and
 to derive the consistent breaking of the symmetry $\zbz\lra \ZBZ$.

Now, in term of the $\cK(2)$ ghost the previous equation \rf{sc2}
rewrites,
\begin{eqnarray}
\sS \cC^{(2)} = \cK^{z}(2)\prt\cC^{(2)}
+2\cC^{(2)}\prt\cK^{z}(2)-\frac{3}{2}
 \beta\cC^{(2)}\prt \cC^{(2)}
\lbl{w22}
\end{eqnarray}
where Eq.\rf{cK} has been used. In particular we get,
\begin{eqnarray}
\sS\biggl(\cC^{(2)}\prt \cC^{(2)}\biggr)=\cK^{z}(2)\prt(\cC^{(2)}\prt
\cC^{(2)})-3 \cC^{(2)}\prt \cC^{(2)}\prt\cK^{z}(2).
\lbl{sc222}
\end{eqnarray}
The nilpotency condition on $\cC^{(2)}$ reads,
\begin{eqnarray}
0 = \sS^2 \cC^{(2)} =
\biggl(\sS\cK^{z}(2)-\cK^{z}(2)\prt\cK^{z}(2)\biggr)
\prt\cC^{(2)}
-2\cC^{(2)}\prt\biggl(\sS\cK^{z}(2)-\cK^{z}(2)\prt\cK^{z}(2)\biggr)-
\nn\\ 
-\ \frac{3}{2}
 \biggl(\sS\beta-\prt(\beta \cK^{z}(2))\biggl) \cC^{(2)}\prt 
\cC^{(2)}.
\lbl{w222}
\end{eqnarray}
Modifying in a consistent way the BRS transformations of both $C$ 
and $\lambda$ while preserving  Eq.\rf{cK}, allows one to set,
\begin{eqnarray}
\sS\cC =\cC\prt\cC + \cX,
\lbl{scx}
\end{eqnarray}
where, from the nilpotency condition,
\begin{eqnarray}
\sS\cX =\cC\prt\cX -\prt\cC\cX.
\lbl{s2}
\end{eqnarray}
Now from $\sS^2\cC^{(2)}=0$ we obtain,
\begin{eqnarray}
\cX \prt\cC^{(2)}= 2\cC^{(2)}\prt \cX
\end{eqnarray}
which can be solved by setting,
\begin{eqnarray}
\cX =\cC^{(2)}\prt \cC^{(2)}\frac{16 }{3}\cT +\alpha\biggl(
\prt\cC^{(2)}\prt^2\cC^{(2)}-
\frac{2}{3}\cC^{(2)}\prt^3\cC^{(2)}\biggr)
\lbl{X0}
\end{eqnarray}
where we have been forced to introduce a spin $(2,0)$-conformal field
$\cT$.
Since $\cX$ has to satisfy Eq.\rf{s2},
$\cT$ is subject to the consistency condition,
\begin{eqnarray}
\sS \cT =\cC\prt\cT +2 \prt\cC \cT+\prt\cC^{(2)}\cW +\frac{2}{3}
\cC^{(2)}
\prt\cW +\alpha\prt^3 \cC 
\lbl{ct1}
\end{eqnarray}
which allows us to introduce a spin $(3,0)$-conformal field $\cW$
whose BRS behaviour can be calculated from the nilpotency
condition applied on \rf{ct1}.
To sum up, we have computed the most general deformation of the $\cC$
ghost field compatible with the fixed variation \rf{sc2} of $\cC^{(2)}$.

From the very definition \rf{cK} of $\cK^z(2)$, and after using
\rf{w22}, the variation $\sS \cK^z(2)$ can be expressed in terms of
both the $\cK^z(2)$ and $\cC^{(2)}$ ghosts only. One gets,
\begin{eqnarray}
 \biggl(\sS \cK^z(2) -\cK^z(2)\prt \cK^z(2)\biggr) -
\biggl(\sS \biggl(\frac{\beta_z}{2}\biggr)-\prt\biggl(\frac{\beta_z}{2}  
\cK^z(2)\biggr)\biggr)  \cC^{(2)}
=\cX - \cC^{(2)}\prt \cC^{(2)}\biggl( \frac{\beta^2_z}{2} \biggr)
\lbl{conditioninv1}
\end{eqnarray}
Requiring the diffeomorphism symmetry implemented by \rf{sK},
on the one hand, Eq.\rf{w222} reduces to, 
\begin{eqnarray}
\biggl(\cC^{(2)}\prt 
\cC^{(2)}\biggr)\biggl[\sS \biggl({\beta_z}\biggr)-
\prt\biggl({\beta_z}\cK^z(2)\biggr)\biggr]=0. 
\lbl{diff22}
\end{eqnarray}
and on the other hand, by plugging \rf{X0} into \rf{conditioninv1},
one gets,
\begin{eqnarray}
\cC^{(2)}
\biggl(\sS \biggl(\frac{\beta_z}{2}\biggr)-\prt\biggl(\frac{\beta_z}{2}  
\cK^z(2)\biggr)\biggr)  
= \cC^{(2)}\prt \cC^{(2)}\biggl(\frac{16 }{3}\cT- \frac{\beta^2_z}{2}
\biggr)
+\alpha\biggl( \prt\cC^{(2)}\prt^2\cC^{(2)}-
\frac{2}{3}\cC^{(2)}\prt^3\cC^{(2)}\biggr).
\lbl{conditioninv2}
\end{eqnarray}
A direct comparison shows that $\alpha=0$, and thus
\begin{eqnarray}
\sS \frac{\beta_z}{2}=\prt\biggl(\frac{\beta_z}{2}  
\cK^z(2)\biggr)
+\prt \cC^{(2)}\biggl(\frac{16 }{3}\cT- \frac{\beta^2_z}{2} \biggr) 
+\Sigma\, \cC^{(2)},
\lbl{conditioninv3}
\end{eqnarray}
where $\Sigma$ is a ghost graded zero quadratic differential which
labels a
$\Phi$-$\Pi$ ambiguity.
The nilpotency allows to compute the BRS variation of
$\Sigma$. However as above, the latter will be exhibited up to
a $\Phi$-$\Pi$ ambiguity and so on. The closure of a minimal BRS
algebra will be achieved by imposing,
\begin{eqnarray}
\frac{16 }{3}\cT= \frac{\beta^2_z}{2}, \qquad
\Sigma=0,
\lbl{conditioninv4}
\end{eqnarray}
so that we are left with
\begin{eqnarray}
\sS \frac{\beta_z}{2}=\prt\biggl(\frac{\beta_z}{2}  
\cK^z(2)\biggr).
\end{eqnarray}
From Eq.\rf{ct1} we derive
\begin{eqnarray}
\cW={\biggl(\frac{3}{2}\biggr)}^\frac{3}{2}{\biggl(\frac{\beta_z}{4}\biggr)}^3 
\lbl{conditioninv5}
\end{eqnarray}
showing that both the fields $\cT$ and $\cW$ are non local fields.
Finally for the sake of consistency, one checks,
\begin{eqnarray}
\sS \biggl(\beta \cK^{z}(2)\biggr)=0.
\end{eqnarray}

In the case \rf{diff22} the smooth change of complex coordinates $\zbz
\lra \ZBZdue$ will be preserved under diffeomorphisms and the ghost
$\cK^z(2)$  
transforms as a factorized ghost vector field, while
\begin{eqnarray}
\sS\lambdaZ=\prt\biggl(\lambdaZ\cK^{z}(2)\biggr).
\end{eqnarray}

The parameter $\lambdaZ$ is compatible with the 
 complex structure defined by the $Z^{(2)}$ coordinates,
and the geometrical quantities $\lambdaZ$ ,$\lambdadue$  
are covariant ; in particular
 the object:
\begin{eqnarray}
\cJ\equiv \frac{\lambdadue}{\lambdaZ}
\end{eqnarray}
 transforms as:
\begin{eqnarray}
\sS\cJ=\cK^{z}(2)\prt\cJ.
\end{eqnarray}
The latter implies,
\begin{eqnarray}
(\bprt-\mu(\zbz,2)\prt)\cJ\equiv\frac{\prt}{\prt {\ovl{Z^{(2)}}}}\cJ=0
\end{eqnarray}
so that $\cJ$ is holomorphic in $Z^{(2)}$. So we can parametrize:
\begin{eqnarray}
\lambda=\frac{\beta}{2\cJ}, \qquad
\lambdadue=\frac{\beta}{2\cJ^2},
\end{eqnarray}

In conclusion the well-defined (chiral) BRS algebra 
(already treated in \cite{Sorella} in a BRS framework), defined by 
\begin{eqnarray}
\sS \cC=\cC\prt\cC+\frac{1}{2}\beta^2\cC^{(2)}\prt \cC^{(2)},\qquad
\sS \cC^{(2)} = \cC\prt\cC^{(2)} +2\cC^{(2)}\prt\cC,\qquad
\sS\beta=\prt\biggl(\beta\biggl(\cC+\frac{1}{2}
\cC^{(2)}\beta\biggr)\biggr)
\lbl{w21}
\end{eqnarray}
describes a diffeomorphism symmetry hidden in the choice of the
 parameters which leads to a diffeomorphism ghost 
$\cK^z(2)=\cC +\frac{\beta_z}{2} \cC^{(2)}$.

The $\beta_z$ parameter describes 
the  relative  tuning  between  the change of complex coordinates
$\ZBZ\lra\ZBZdue$ with respect to the same $\zbz$ background, namely
\begin{eqnarray}
\beta_z= \frac{\prt\cK^z(2)}{\prt\cC^{(2)}}.
\lbl{beta}
\end{eqnarray}

\subsect{The induced $W_3$-gravity}

\indent

According to Theorem \ref{thm}, one considers as a canonical
transformation the following
vertical smooth change of coordinates $Z\zY \lra Z'\zY$, 
where $Z\zY=\prt_{\YZ}\Phi(z,Y)$ and where 
$Z'\zY$ is defined by the following replacement,
\begin{eqnarray}
Z^{(2)}\zbz\longrightarrow
Z^{(2)}(\zbz,\YZ)\equiv\frac{1}{2}{\biggl(\frac{\prt}{\prt\YZ}\biggr)}^2
\Phi\zY = \sum_{n\geq 0} \frac{(n+2)(n+1)}{2} \YZ^n \,Z^{(2+n)}\zbz
\lbl{zeta2}
\end{eqnarray}
in the expansion \rf{Zseries} of the $Z\zY$ coordinate. One has
\begin{eqnarray}
Z'\zY = Z\zbz + 2\,\YZ Z^{(2)}\zbz + \sum_{n\geq 2} (n+1)^2\, \YZ^n\,
Z^{(n+1)}\zbz ,
\end{eqnarray}
showing that the corresponding generating function $\Phi'$ coincides
with $\Phi$ up to the second order in $\YZ$.
Furthermore, $Z^{(2)}(\zbz,\YZ)$ has as BRS variation,
\begin{eqnarray}
\sS Z^{(2)}(\zbz,\YZ)=
\frac{1}{2}{\biggl(\frac{\prt}{\prt\YZ}\biggr)}^2 \Lambda\zY
\equiv\frac{1}{2} \sum_{n\geq 0} (n+2)(n+1)\YZ^n 
\Upsilon^{(n+2)}\zbz = \nn\\
=\lambdadue\zbz
\cC\zbz+\lambda^2\zbz\cC^{(2)}\zbz+\Sigma^{Z^{(2)}}(\zbz,\YZ)
\equiv\lambdadue(\zbz,\YZ){\cK'}^z_2(\zbz,\YZ),
\end{eqnarray}
where in complete analogy with the construction given in section 3 we
have,
\begin{eqnarray}
\sS{\cK'}^z_2(\zbz,\YZ)={\cK'}^z_2(\zbz,\YZ)\prt{\cK'}^z_2(\zbz,\YZ)
\lbl{sc22}
\end{eqnarray}
Now it is easy to realize that
\begin{eqnarray}
\lambdadue(\zbz,\YZ)\equiv\prt  Z^{(2)}(\zbz,\YZ)=\lambdadue\zbz+
\sum_{n\geq 1} (n+2)(n+1)\YZ^n \,\lambda^{Z^{(2+n)}}_z\zbz
\end{eqnarray}
from which, similarly to the previous example, it follows that, 
\begin{eqnarray}
{\cK'}^z_2(\zbz,\YZ)=\cC\zbz+\frac{\lambda^2\zbz}
{\lambdadue(\zbz,\YZ)}\cC^{(2)}\zbz+\omega^z(\zbz,\YZ)\nn\\
\equiv \cC\zbz +\frac{\beta'_z(\zbz,\YZ)}{2}\cC^{(2)}\zbz+
\omega^z(\zbz,\YZ)
\end{eqnarray}
Therefore if we impose the condition Eq\rf{sc22} together with 
Eqs.\rf{scx}\rf{sc2}, the previous condition $\alpha =0$
 can be avoided since the B.R.S variations of $\beta'_z(\zbz,\YZ)$ and
$\omega^z(\zbz,\YZ)$ 
are at our disposal for this purpose.
We do not write these variations since 
they are unimportant in the treatment.
But it has to be noted that 
in the $\zbz$ plane (with no $\YZ$ dependence) only the 
couple of ghost fields $\cC\zbz$, $\cC^{(2)}\zbz$ will survive and
whose BRS variations are given by,
\begin{eqnarray}
\sS \cC &=& \cC\prt\cC
- \cC^{(2)}\prt \cC^{(2)} \frac{16 }{3}\cT +\ \alpha\biggl( 
\prt\cC^{(2)}\prt^2\cC^{(2)}
- \frac{2}{3}\cC^{(2)}\prt^3\cC^{(2)}\biggr)\nn \\
\sS \cC^{(2)} &=& \cC\prt\cC^{(2)} +2\cC^{(2)}\prt\cC,
\end{eqnarray}
where for the sake of definiteness $\cT$ turns out to be a projective
connection, in contrast to the 
quadratic differential constructed in the previous example.
The nilpotency condition infers altogether that $\cT$ is assigned 
a well defined transformation law 
\begin{eqnarray}
\sS \cT =\cC\prt\cT +2 \prt\cC \cT-\prt\cC^{(2)}\cW -\frac{2}{3}
\cC^{(2)}
\prt\cW +\alpha\prt^3 \cC ,
\end{eqnarray}
in terms of a cubic differential $\cW$, whose variation is,
\begin{eqnarray}
\sS \cW &=& \cC \prt \cW + 3\prt\cC \cW + \frac{2}{3} \cT
\prt\biggl(\cC^{(2)}\cT\biggr)+\frac{\alpha}{24}\biggl(\alpha\prt^5\cC^{(2)}
+2\cC^{(2)}\prt^3\cT \nn\\ 
&&\qquad +\ 10\cT\prt^3\cC^{(2)}+15\prt\cT\prt^2\cC^{(2)}+
9\prt^2\cT\prt\cC^{(2)}\biggr).
\end{eqnarray}
However the nilpotency condition on $\cW$ holds whatever $\alpha$ is.
Requiring that all the variations are globally defined
infers that $\alpha=1$. Thus we get the well-defined exact differential
system,
\begin{eqnarray}
\sS \cC &=&\cC\prt\cC -\frac{16}{3}
\cC^{(2)}\prt \cC^{(2)}\cT +\biggl( \prt\cC^{(2)}\prt^2\cC^{(2)}-
\frac{2}{3}\cC^{(2)}\prt^3\cC^{(2)}\biggr)\nn \\
\sS \cC^{(2)} &=& \cC\prt\cC^{(2)} +2\cC^{(2)}\prt\cC \nn \\
\sS \cT &=& \cC\prt\cT +2 \prt\cC \cT -24\prt\cC^{(2)}\cW -16 \cC^{(2)}
\prt\cW +\prt^3 \cC \nn \\
\sS \cW &=& \cC \prt \cW + 3\prt\cC \cW + \frac{2}{3}
\cT \prt\biggl(\cC^{(2)}\cT\biggr)+\frac{1}{24}\biggl(\prt^5\cC^{(2)}\nn
\\
&&\qquad +\ 2\cC^{(2)}\prt^3\cT
+10\cT\prt^3\cC^{(2)}+15\prt\cT\prt^2\cC^{(2)}+
9\prt^2\cT\prt\cC^{(2)}\biggr)
\end{eqnarray}
This BRS algebra has already been described in a different context by
\cite{Grimm}, \cite{Ader1}.

\sect{Conclusions}

\indent

In this paper we have introduced a systematic approach to $W$-algebras 
based on canonical transformations
by means of an abstract symplectic structure. It is however as claimed
by Witten and Hull, the action of diffeomorphisms of the cotangent
bundle which generates these $W$-symmetries.
Our construction provides a BRS formulation of a local symmetry
which is recognized to be that of $W$-algebras (see Bilal  et al. in
\cite{geom}, and \cite{Zucchini,Sorella,many,OSSN,Grimm,Ader1,Ader2})
but without any care about the OPE's  
of primary fields. It may also shed some new light in the study 
of their geometrical structure \cite{geom}.
 
As shown for the $W_3$ specific examples, various $W$-algebras may
emerge from the construction.
Moreover, one may ask oneself about the link between the 
truncation procedure in the $\YZ$ variable
(which fixes  the ``rank" of the algebra) and the relative conformal weights 
of the primary fields whose OPE's give rise to the algebras.
One answer to this question might originate from the locality
principle of QFT and the intrinsic symplectic structure introduced 
through the canonical quantization scheme. Indeed, by fixing the
maximum order  
of the coframe  in the $\YZ$ coordinate, the relative ``field momentum 
representation" in the associated Field theory carries this 
operation in its tensorial content, but the mechanism is not 
completely understood by the authors yet.
In the paper, we have also presented a possible space-time origin of
these algebras which ought to be useful in the study of induced $W$-gravity.
Although most of our discussion have been rather complete and
general, the 
absence of a physical-phenomenological framework is the more 
important limit of the present treatment. 
  
More examples are needed for further investigations in the role
played by such symmetries in physics, as recalled in the introduction.
This requires either a Lagrangian 
or an Hamiltonian approach in which the quantization procedure 
can be performed according to the physical context. The former
 will be studied elsewhere.

\indent

\noindent
{\bf Acknowledgements.} We are grateful to A.~Blasi and N.~Maggiore for
helpful discussions and comments on the manuscript.

\end{document}